\documentclass[twocolumn,showpacs,showkeys,amsmath,amssymb,superscriptaddress,floatfix,nofootinbib]{revtex4}

\usepackage{graphicx}
\usepackage{bm} 
\usepackage{subfigure}

\def\vec#1{\mathchoice{\mbox{\boldmath$\displaystyle#1$}}
{\mbox{\boldmath$\textstyle#1$}}
{\mbox{\boldmath$\scriptstyle#1$}}
{\mbox{\boldmath$\scriptscriptstyle#1$}}}
\makeatletter
\newcommand\erfc{\mathop{\operator@font erfc}\nolimits}
\def\slashchar#1{\setbox0=\hbox{$#1$}
   \dimen0=\wd0 \setbox1=\hbox{/} \dimen1=\wd1
   \ifdim\dimen0>\dimen1 \rlap{\hbox to \dimen0{\hfil/\hfil}} #1
   \else  \rlap{\hbox to \dimen1{\hfil$#1$\hfil}} / \fi}

\begin{document}
 
\title{
Highly-anisotropic and strongly-dissipative hydrodynamics for early stages of relativistic heavy-ion collisions
\footnote{Supported in part by the Polish Ministry of Science and Higher Education, grant  N202 034 32/0918.}}

\author{Wojciech Florkowski} 
\email{Wojciech.Florkowski@ifj.edu.pl}
\affiliation{Institute of Physics, Jan Kochanowski University, PL-25406~Kielce, Poland} 
\affiliation{The H. Niewodnicza\'nski Institute of Nuclear Physics, Polish Academy of Sciences, PL-31342 Krak\'ow, Poland}

\author{Radoslaw Ryblewski} 
\email{Radoslaw.Ryblewski@ifj.edu.pl}
\affiliation{The H. Niewodnicza\'nski Institute of Nuclear Physics, Polish Academy of Sciences, PL-31342 Krak\'ow, Poland}

\date{January 26, 2008}

\begin{abstract}
We introduce a new framework of highly-anisotropic hydrodynamics that includes dissipation effects. Dissipation is defined by the form of the entropy source that depends on the pressure anisotropy and vanishes for the isotropic fluid. With a simple ansatz for the entropy source obeying general physical requirements, we are led to a non-linear equation describing the time evolution of the anisotropy in purely-longitudinal boost-invariant systems. Matter that is initially highly anisotropic approaches naturally the regime of the perfect fluid. Thus, the resulting evolution agrees with the expectations about the behavior of matter produced at the early stages of relativistic heavy-ion collisions. The equilibration is identified with the processes of entropy production.
\end{abstract}



\maketitle 

\section{Introduction}
\label{sect:intro}

The experimental results obtained in the heavy-ion experiments at the Relativistic Heavy-Ion Collider (RHIC), in particular the large values of the elliptic flow, are most often interpreted as the evidence for a very fast equilibration of the produced matter (presumably within a fraction of 1~fm/c) and for its almost perfect-fluid behavior  \cite{Kolb:2003dz,Huovinen:2003fa,Shuryak:2004cy,Teaney:2001av,Hama:2005dz,Hirano:2007xd,Nonaka:2006yn}. 

The very fast equilibration is naturally explained within a concept that the produced matter is a strongly coupled quark-gluon plasma (sQGP) \cite{Shuryak:2004kh}. However, there exist other explanations that assume that the plasma is weakly interacting. In this case the plasma instabilities lead to the fast isotropization of matter, which in turn helps to achieve the full equilibration in a short time \cite{Mrowczynski:2005ki}. 

Recently, several explicit calculations have shown that the large values of the elliptic flow and other soft-hadronic observables may be successfully reproduced in the models that do not assume the very fast equilibration. For example, in Refs. \cite{Broniowski:2008qk} the stage described by the perfect-fluid hydrodynamics was preceded by the free streaming of partons (see also \cite{Gyulassy:2007zz,Akkelin:2009nz}), while in Refs. \cite{Bialas:2007gn,Chojnacki:2007fi} the authors assumed that only transverse degrees of freedom are thermalized (schematic scenarios describing the approach towards the full equilibration were discussed in this context in Refs. \cite{Ryblewski:2009hm,Ryblewski:2010tn}). Such results indicate that the assumption of the fast equilibration/isotropization might be relaxed. 

It should be also emphasized that the concept of practically instantaneous equilibration seems to contradict the results of the microscopic models of heavy-ion collisions. Such models typically use the ideas of color strings or color-flux tubes. The system produced by strings is highly anisotropic; the pressure in the direction transverse to the collision axis is usually much larger than the longitudinal pressure\footnote{As usual, the longitudinal direction is defined by the direction of the beam.}. 

Similar situation takes place in the Color Glass Condensate (CGC) approach where the distribution functions are far away from the equilibrium ones. In this case the longitudinal momentum distribution is much narrower than the transverse one and described by the Dirac delta function $\delta(p_\parallel)$ at $z=0$ \cite{Kovner:1995ja,Bjoraker:2000cf}. This approximation is often used in descriptions of the initial stage in nucleus-nucleus collisions (for example see \cite{El:2007vg}).

In view of the problems connected with the equilibration and isotropization of the plasma, it is useful to develop and analyze the models which can be used to describe locally anisotropic systems. In this paper we introduce the framework of highly-anisotropic hydrodynamics that takes into account dissipation effects. The dissipation is defined by the form of the entropy source. The latter depends on the pressure anisotropy and vanishes for the isotropic fluid. The proposed model has a structure that is very much similar to the perfect-fluid hydrodynamics. The main two differences are connected with: {\bf i)} the possibility that the longitudinal and transverse pressures are different, and {\bf ii)} the possibility of entropy production. By relaxing the assumption about the isentropic flow, we generalize our previous formulations of anisotropic (magneto)hydrodynamics presented in \cite{Florkowski:2008ag,Florkowski:2009sw}. 

It is important to note that the deviations from equilibrium are naturally described in the framework of viscous (Israel-Stewart) hydrodynamics. However, the region of the applicability of viscous hydrodynamics extends to the systems that are close to equilibrium. This is reflected in the dependence of the transport coefficients on the equilibrium variables such as temperature or chemical potentials. Thus, the viscous corrections are applicable for the intermediate, locally almost equilibrated stage (for a recent review see \cite{Heinz:2009xj}).  

In our opinion, the use of viscous hydrodynamics in the description of very early stages of the collisions may be inadequate --- the strong reduction of the initial longitudinal pressure leads to significant deviations from equilibrium (see \cite{Bozek:2007di}). On the other hand, the kinetic models that are most suitable for the description of the systems out of equilibrium are very much complicated and difficult to deal with. Therefore, there is a place for effective models which can describe the early non-equilibrium dynamics together with the transition to the perfect-fluid regime. Our formulation of the anisotropic hydrodynamics follows this direction.

Within our approach, a simple ansatz for the entropy source leads to a non-linear equation describing the time evolution of the anisotropy in the purely-longitudinal boost-invariant systems. The non-linearity implies that a possible strong initial anisotropy is eliminated. The resulting evolution of the system agrees with the expectations about the behavior of matter produced at the early stages of relativistic heavy-ion collisions. In particular, the equilibration of the system is connected with the processes of entropy production.

Although our numerical results are presented for the simple one-dimensional system, the proposed formalism is general and may be applied to more complicated 2+1 and 3+1 situations (in a similar way as the perfect-fluid hydrodynamics). In addition, different forms of the entropy source inspired by different microscopic mechanisms may be analyzed. In our further studies we want to explore such rich possibilities.

\medskip

Below we assume that particles (partons) are massless and we use the following definitions for rapidity and spacetime rapidity,
\begin{eqnarray}
y = \frac{1}{2} \ln \frac{E_p+p_\parallel}{E_p-p_\parallel}, \quad
\eta = \frac{1}{2} \ln \frac{t+z}{t-z}, \label{yandeta} 
\end{eqnarray}
which come from the standard parameterization of the four-momentum and spacetime coordinate of a particle,
\begin{eqnarray}
p^\mu &=& \left(E_p, {\vec p}_\perp, p_\parallel \right) =
\left(p_\perp \cosh y, {\vec p}_\perp, p_\perp \sinh y \right), \nonumber \\
x^\mu &=& \left( t, {\vec x}_\perp, z \right) =
\left(\tau \cosh \eta, {\vec x}_\perp, \tau \sinh \eta \right). \label{pandx}
\end{eqnarray} 
Here the quantity $p_\perp$ is the transverse momentum
\begin{equation}
p_\perp = \sqrt{p_x^2 + p_y^2},
\label{energy}
\end{equation}
and $\tau$ is the (longitudinal) proper time
\begin{equation}
\tau = \sqrt{t^2 - z^2}.
\label{tau}
\end{equation} 
Throughout the paper we use the natural units where $c=1$ and $\hbar=1$.

\section{Anisotropic hydrodynamics with dissipation}
\label{sect:aniso-system}

\subsection{Energy-momentum tensor and entropy flux}
\label{sect:TmunuSmu}

Our approach is based on the following form of the energy-momentum tensor
\begin{eqnarray}
T^{\mu \nu} &=& \left( \varepsilon  + P_\perp\right) U^{\mu}U^{\nu} 
- P_\perp \, g^{\mu\nu} - (P_\perp - P_\parallel) V^{\mu}V^{\nu}, \nonumber \\
\label{Tmunudec}
\end{eqnarray}
where $\varepsilon$, $P_\perp$, and $P_\parallel$ are the energy density, transverse pressure, and longitudinal pressure, respectively. In the special case of the isotropic fluid, where \mbox{$P_\perp=P_\parallel=P$}, we recover the form of the energy-momentum tensor of the perfect-fluid hydrodynamics. The four-vector $U^\mu$ describes the hydrodynamic flow, 
\begin{equation}
U^\mu = \gamma (1, {\bf v}), \quad \gamma = (1-v^2)^{-1/2},
\label{Umu}
\end{equation}
while $V^\mu$ defines the direction of the longitudinal axis that plays a special role due to the initial geometry of the collision. The four-vectors $U^\mu$ and $V^\mu$ satisfy the following normalization conditions
\begin{eqnarray}
U^2 = 1, \quad V^2 = -1, \quad U \cdot V = 0.
\label{UVnorm}
\end{eqnarray}
In the local-rest-frame (LRF) of the fluid element we have $U^\mu = (1,0,0,0)$. In the same reference frame the four-vector $V^\mu$ that satisfies Eqs. (\ref{UVnorm}) and defines the longitudinal direction has the form $V^\mu = (0,0,0,1)$. The forms of $U^\mu$ and $V^\mu$ in other reference systems are obtained by the Lorentz boosts.

In LRF the energy-momentum tensor has the diagonal structure,
\begin{equation}
T^{\mu \nu} =  \left(
\begin{array}{cccc}
\varepsilon & 0 & 0 & 0 \\
0 & P_\perp & 0 & 0 \\
0 & 0 & P_\perp & 0 \\
0 & 0 & 0 & P_\parallel
\end{array} \right).
\label{Tmunuarray}
\end{equation}
Hence, as expected, the formula (\ref{Tmunudec}) allows for different pressures in the longitudinal and transverse directions.

In addition to the energy-momentum tensor (\ref{Tmunudec}) we introduce the entropy flux
\begin{eqnarray}
\sigma^{\mu} &=& \sigma U^{\mu},
\label{smudec}
\end{eqnarray}
where $\sigma$ is the entropy density. We assume that $\varepsilon$ and  $\sigma$ are functions of $P_\perp$ and $P_\parallel$. In particular, since we consider massless partons here, the condition $T^\mu_{\,\,\,\mu}=0$ gives
\begin{equation}
\varepsilon = 2 P_\perp + P_\parallel.
\label{eos}
\end{equation}

We note that the form of the energy-momentum tensor (\ref{Tmunudec}) resembles the form used in relativistic magnetohydrodynamics \cite{PhysRevE.47.4354,PhysRevE.51.4901}. In that case the anisotropy is induced by the presence of the magnetic field. At the early stages of heavy-ion collisions we have similar situation --- there exist strong color magnetic and electric longitudinal fields (Glasma following CGC \cite{Lappi:2006fp}) which polarize the medium. The {\it explicit} inclusion of the fields should be one of the first tasks connected with the generalization of the presented framework. The first steps in this direction but for the isentropic case were made in \cite{Florkowski:2009sw}.

\subsection{Evolution equations}
\label{sect:evol}

The dynamics of the system is governed by the equations expressing the energy-momentum conservation and the entropy growth (the second law of thermodynamics),
\begin{eqnarray}
\partial_\mu T^{\mu \nu} &=& 0, \label{enmomcon} \\
\partial_\mu \sigma^{\mu} &=& \Sigma. \label{engrow}
\end{eqnarray}
Here the function $\Sigma$ describes the entropy source. The form of $\Sigma$ must be treated as the assumption defining the dynamics of the anisotropic fluid. Besides the condition $\Sigma \geq 0$ it is natural to assume that \mbox{$\Sigma = 0$} for \mbox{$P_\perp=P_\parallel$}. In this way, in the case where the two pressures are equal, we recover the structure of the perfect-fluid hydrodynamics. We note that $\Sigma$ is the {\it internal} source of the entropy, i.e., $\Sigma$ describes the entropy growth due to the equilibration of pressures in the system. 

In the following we shall treat $\Sigma$ as a function of $P_\perp$ and $P_\parallel$. In this way, Eqs. (\ref{enmomcon}) and (\ref{engrow}) form a closed system of 5 equations for 5 unknown functions: three components of the fluid velocity, $P_\perp$, and $P_\parallel$. The projections of Eq. (\ref{enmomcon}) on $U_\nu$ and $V_\nu$ gives

\begin{eqnarray}
U^\mu \partial_\mu \varepsilon &=& - \left( \varepsilon+P_\perp \right) \partial_\mu U^\mu 
+ \left( P_\perp-P_\parallel \right) U_\nu  V^\mu \partial_\mu V^\nu, \nonumber \\
&& \label{enmomconU} \\
V^\mu \partial_\mu P_\parallel &=& - \left( P_\parallel-P_\perp \right) \partial_\mu V^\mu 
+ \left( \varepsilon+ P_\perp \right) V_\nu  U^\mu \partial_\mu U^\nu. \nonumber \\
&& \label{enmomconV}
\end{eqnarray}

\subsection{Anisotropic momentum distribution}
\label{sect:aniso-distribution}

In our previous paper \cite{Florkowski:2009sw} we showed that the structure (\ref{Tmunudec})--(\ref{smudec}) follows from the following form of the distribution function
\begin{equation}
f = f\left( \frac{p_\perp}{\lambda_\perp},\frac{|p_\parallel|}{\lambda_\parallel}\right).
\label{Fxp1}
\end{equation}
Here the two parameters $\lambda_\perp$ and $\lambda_\parallel$ may be interpreted as the transverse and longitudinal temperature. The form (\ref{Fxp1}) is valid in the local rest frame of the fluid where $U^\mu = (1,0,0,0)$ and $V^\mu=(0,0,0,1)$. The explicitly covariant form of the distribution function (\ref{Fxp1}) has the form
\begin{equation}
f  = f\left( \frac{\sqrt{(p \cdot U)^2 - (p \cdot V)^2 }}{\lambda _\perp }, 
\frac{|p \cdot V|}{\lambda _\parallel  }\right).
\label{Fxp2}
\end{equation}

In this paper we consider the exponential distribution function which in the local rest frame has the form
\begin{equation}
f = g_0 \exp \left( -\sqrt{\frac{p_\perp ^2}{\lambda_\perp  ^2} + 
\frac{p_\parallel^2}{\lambda_\parallel^2} } \,  \right).
\label{Boltz1}
\end{equation}
Equation (\ref{Boltz1}) may be regarded as the generalization of the Boltzmann equilibrium distribution where \mbox{$\lambda_\perp = \lambda_\parallel = T$}.  The parameter $g_0$ is the degeneracy factor connected with internal quantum numbers. Having in mind the fact that the initially produced matter consists mainly of gluons we obtain 
\begin{equation}
g_0 = 16.
\label{g0}
\end{equation}

Using the {\it covariant} form of (\ref{Boltz1}) in the definition of the energy-momentum tensor,
\begin{eqnarray}
T^{\mu \nu} &=& \int \frac{d^3p}{(2\pi)^3 \, E_p} \, p^{\mu} p^\nu f,
\label{TmunuKin} 
\end{eqnarray}
and in the definition of the entropy flux\footnote{The formula (\ref{SmuKin}) assumes the classical Boltzmann statistics. It may be generalized to the case of bosons or fermions in the standard way.}
\begin{equation}
\sigma^\mu =  \int \frac{d^3p}{(2 \pi)^3} \frac{p^\mu}{E_p} f
 \, \left[1 -  \ln \left(\frac{f}{g_0}\right) \right],
\label{SmuKin}
\end{equation}
we obtain Eqs. (\ref{Tmunudec}) and (\ref{smudec}).

The energy density, transverse pressure, longitudinal pressure, and entropy density are obtained from the following integrals
\begin{eqnarray}
\varepsilon &=&   \int \frac{d^3p}{ (2\pi)^3}  \, E_p \, f\left( \frac{p_\perp}{\lambda_\perp}, 
\frac{| p_\parallel |}{\lambda_\parallel }\right),  \label{epsilon1} 
\end{eqnarray}
\begin{eqnarray}
P_\perp &=&   \int \frac{d^3p}{ (2\pi)^3}  \, \frac{p_\perp^2}{2 E_p} \, f\left( \frac{p_\perp}{\lambda _\perp}, \frac{| p_\parallel |}{\lambda _\parallel }\right),  \label{PT1} 
\end{eqnarray}
\begin{eqnarray}
P_\parallel &=&   \int \frac{d^3p}{ (2\pi)^3}  \, \frac{p_\parallel^2}{E_p} \, f\left( \frac{p_\perp}{\lambda _\perp}, \frac{| p_\parallel |}{\lambda _\parallel }\right),  \label{PL1} 
\end{eqnarray}
\begin{equation}
\sigma =  \int \frac{d^3p}{(2 \pi)^3} f
 \, \left[1 -  \ln \left(\frac{f}{g_0}\right) \right].
\label{S}
\end{equation}

\begin{figure}[t]
\begin{center}
\subfigure{\includegraphics[angle=0,width=0.45\textwidth]{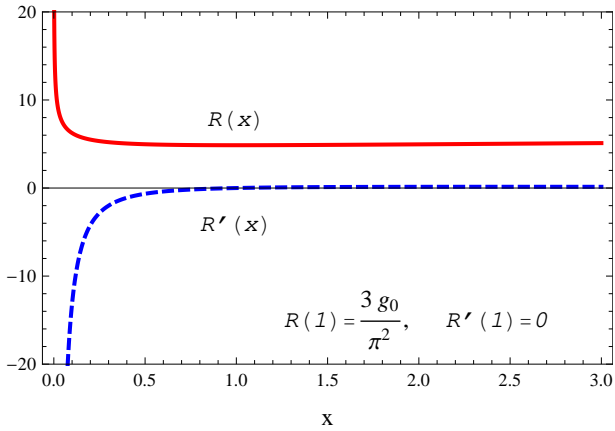}} \\
\end{center}
\caption{(Color online) The $x$-dependence of the function $R(x)$ and its derivative $R'(x)$. }
\label{fig:R}
\end{figure}

\subsection{Pressure anisotropy}
\label{sect:temp-aniso}

Equations (\ref{epsilon1})--(\ref{S}) allow us to express all thermodynamic quantities\footnote{We continue to call $\varepsilon$, $P_\perp$, $P_\parallel$, and $\sigma$ the thermodynamic quantities although, strictly speaking, these quantities do not refer to the equilibrium state. Similarly, we call $\lambda_\perp$ and $\lambda_\parallel$ the transverse and longitudinal temperature. The reason for this terminology is close similarity to the equilibrium variables.} in terms of  $\lambda_\perp$ and $\lambda_\parallel$. Thus, instead of $P_\perp$ and $P_\parallel$ we may switch to $\lambda_\perp$ and $\lambda_\parallel$. It turns out, however, that the most useful two independent parameters are the entropy density $\sigma$ and the variable $x$ defined by the expression
\begin{equation}
x = \left( \frac{\lambda_\perp}{\lambda_\parallel} \right)^2.
\label{x}
\end{equation}
Treating $\sigma$ and $x$ as the two independent thermodynamic variables (instead of $P_\perp$ and $P_\parallel$ or instead of $\lambda_\perp$ and $\lambda_\parallel$) we obtain the compact expressions:
\begin{eqnarray}
\varepsilon &=&  \left(\frac{\pi^2 \sigma}{4 g_0} \right)^{4/3} R(x),
\label{epsilon2} 
\end{eqnarray}
\begin{eqnarray}
P_\perp &=&  \left(\frac{\pi^2 \sigma}{4 g_0} \right)^{4/3}
\left[\frac{R(x)}{3} + x R^\prime(x) \right],   
\label{PT2} 
\end{eqnarray}
\begin{eqnarray}
P_\parallel &=&  \left(\frac{\pi^2 \sigma}{4 g_0} \right)^{4/3} 
\left[\frac{R(x)}{3} - 2 x R^\prime(x) \right],
\label{PL2} 
\end{eqnarray}
where the function $R(x)$ is defined by the formula \cite{Florkowski:2009sw} \footnote{Note that for $x < 1$ the function $(\arctan\sqrt{x-1})/\sqrt{x-1}$ should be replaced by $(\hbox{arctanh}\sqrt{1-x})/\sqrt{1-x}$}
\begin{equation}
R(x) = \frac{3\, g_0\, x^{-\frac{1}{3}}}{2 \pi^2} \left[ 1 + \frac{x \arctan\sqrt{x-1}}{\sqrt{x-1}}\right].
\end{equation}
The symbol $R'(x)$ denotes the derivative of $R(x)$ with respect to $x$. For $x=1$ we find $R'(x)=0$ and, as expected, $P_\perp=P_\parallel$.  The $x$-dependence of the function $R(x)$ and its derivative $R'(x)$ is shown in Fig. \ref{fig:R}.

\begin{figure}[t]
\begin{center}
\subfigure{\includegraphics[angle=0,width=0.45\textwidth]{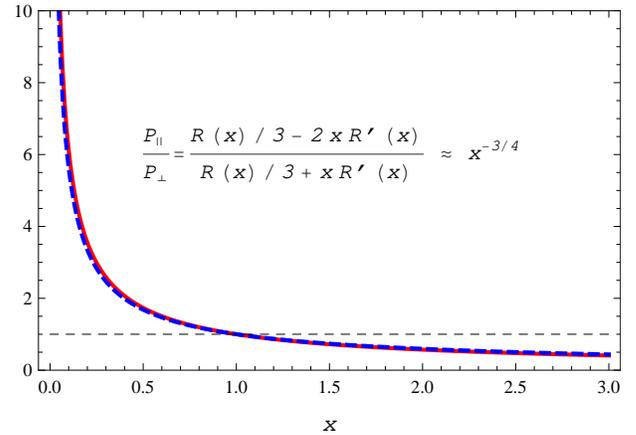}} \\
\end{center}
\caption{(Color online) The $x$-dependence of the ratio of the longitudinal and transverse pressure (thick solid line) and its approximation with the function $x^{-3/4}$ (thick dashed line). }
\label{fig:PA}
\end{figure}

In Fig. \ref{fig:PA} we show the ratio of the longitudinal and transverse pressure plotted as a function of $x$ (solid thick line). The $P_\parallel/P_\perp$ ratio is determined by the $x$-dependence of the function $R(x)$ and its derivative. To a good approximation one finds
\begin{equation}
\frac{P_\parallel}{P_\perp} \approx x^{-3/4}.
\end{equation}
Thus $x$ may be treated as the direct measure of the pressure anisotropy. We note that realistic initial conditions in heavy-ion collisions give $P_\parallel/P_\perp \ll 1$, which corresponds to $x \gg 1$.

\section{Purely-longitudinal boost-invariant motion}
\label{sect:plbimot}

\subsection{Implementation of boost-invariance}
\label{sect:implement}

For purely longitudinal and boost-invariant motion we may write
\begin{equation}
U^{\mu} = (\cosh\eta,0,0,\sinh\eta),
\label{Uv}
\end{equation}
and
\begin{equation}
V^{\mu} = (\sinh\eta,0,0,\cosh\eta).
\label{Vv}
\end{equation}
The boost-invariant character of Eq. (\ref{Fxp2}) is immediately seen if we write the explicit expression for $p \cdot U$ and $p \cdot V$,
\begin{eqnarray}
p \cdot U &=& p_\perp  \cosh(y-\eta), \nonumber \\
p \cdot V &=& p_\perp  \sinh(y-\eta).
\label{pdotUV}
\end{eqnarray}

We also obtain
\begin{eqnarray}
U^\mu \partial_\mu &=& \frac{\partial}{\partial \tau}, \nonumber \\
V^\mu \partial_\mu &=& \frac{\partial}{\tau \partial \eta},
\label{binv1}
\end{eqnarray}
which leads to the equations
\begin{eqnarray}
U^\mu \partial_\mu U^\nu &=& 0,  \nonumber \\
\tau V^\mu \partial_\mu V^\nu &=&  U^\nu,  \nonumber \\
\partial_\mu V^\mu &=& 0.
\label{binv2}
\end{eqnarray}
We note that the boost-invariance requires that all scalar quantities such as $\varepsilon$, $P_T$, or $P_L$ do not depend on space-time rapidity $\eta$.

\subsection{Boost-invariant equations of motion}
\label{sect:bieqmot}

In the considered case, the energy-momentum conservation law (\ref{enmomconU}) is reduced to the equation
\begin{equation}
\frac{d\varepsilon}{d\tau} = -\frac{\varepsilon + P_\parallel}{\tau},
\label{eq0}
\end{equation}
while the entropy conservation yields
\begin{equation}
\frac{d\sigma}{\sigma d\tau} + \frac{1}{\tau} = \frac{\Sigma}{\sigma}.
\label{eq2}
\end{equation}
Equation (\ref{enmomconV}) is automatically fulfilled if the thermodynamic variables do not depend on $\eta$.

By changing to the $x$ variable we may rewrite Eq. (\ref{eq0}) in the form
\begin{equation}
R'(x) \left( \frac{dx}{d\tau} - \frac{2 x}{\tau} \right) =  - \frac{4}{3} R(x) 
\left( \frac{d \sigma}{\sigma d\tau} + \frac{1}{\tau} \right).
\label{eq1}
\end{equation}
Before we proceed further with the analysis of the dissipative flow where $\Sigma >0$, it is useful to consider the non-dissipative flow where $\Sigma=0$. In this case, from Eq. (\ref{eq2}) we recover the Bjorken solution $\sigma = \sigma_0 \tau_0/\tau$ and the right-hand-side of Eq. (\ref{eq1}) vanishes. This implies that either $x=1$ (in which case $R'(1)=0$) or $x = x_0 \tau^2/\tau_0^2$ (in which case $dx/d\tau=2x/\tau$). The parameters $\sigma_0$, $\tau_0$, and $x_0$ are arbitrary constants here. The case $x=1$ corresponds to the standard, perfect-fluid hydrodynamics. The case where $x = x_0 \tau^2/\tau_0^2$ was examined in \cite{Florkowski:2008ag,Florkowski:2009sw}.

\subsection{Ansatz for $\Sigma$}
\label{sect:ansatz}

Equations (\ref{eq2}) and (\ref{eq1}) may be solved only if the dependence of the function $\Sigma$ on the variables $\sigma$ and $x$ is given. The functional form $\Sigma(\sigma,x)$ must be delivered as the external input for our calculations. 

The simplest ansatz for $\Sigma$ that has the correct dimension and satisfies the general conditions that $\Sigma \geq 0$ and $\Sigma(\sigma,x=1)=0$ has the form
\begin{equation}
\Sigma = \frac{(\lambda_\perp-\lambda_\parallel)^2}{\lambda_\perp \,\lambda_\parallel}  
\frac{\sigma}{\tau_{\rm eq}}
= \frac{(1-\sqrt{x})^2}{\sqrt{x}} \frac{\sigma}{\tau_{\rm eq}}.
\label{ansatz1}
\end{equation} 
Here ${\tau_{\rm eq}}$ is a timescale parameter. The natural feature of (\ref{ansatz1}) is the fact that $\Sigma$ is proportional to $\sigma$, hence (\ref{ansatz1}) does not destroy the scale invariance of the perfect-fluid hydrodynamics, that allows for multiplication of $\sigma$ in the evolution equations by an arbitrary constant. Moreover, $\Sigma$ defined by (\ref{ansatz1}) stays constant if $\lambda_\perp$ and $\lambda_\parallel$ are interchanged.

Substituting (\ref{ansatz1}) in (\ref{eq1}) leads to the ordinary differential equation for $x$ only, 
\begin{equation}
 \frac{dx}{d\tau}  =  \frac{2 x}{\tau} - \frac{4 H(x)}{3\tau_{\rm eq}} ,
\label{eq12}
\end{equation}
where we have defined
\begin{equation}
H(x) =  \frac{R(x)}{R'(x)} \frac{(1-\sqrt{x})^2}{\sqrt{x}}.
\label{H}
\end{equation}

\begin{figure}[t]
\begin{center}
\subfigure{\includegraphics[angle=0,width=0.45\textwidth]{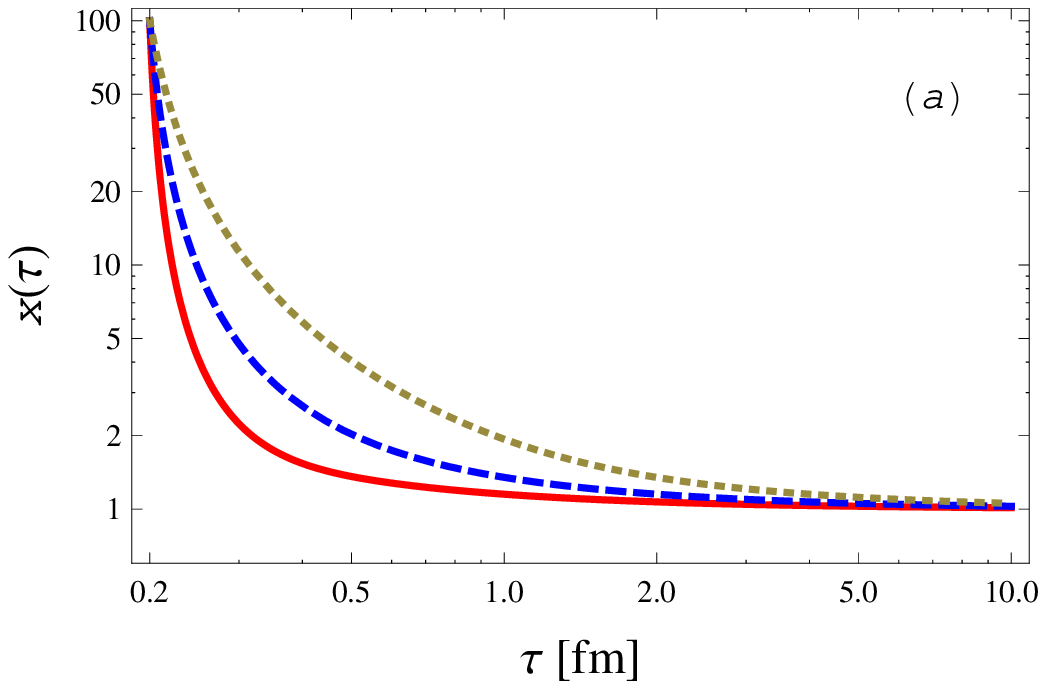}}  \\
\subfigure{\includegraphics[angle=0,width=0.45\textwidth]{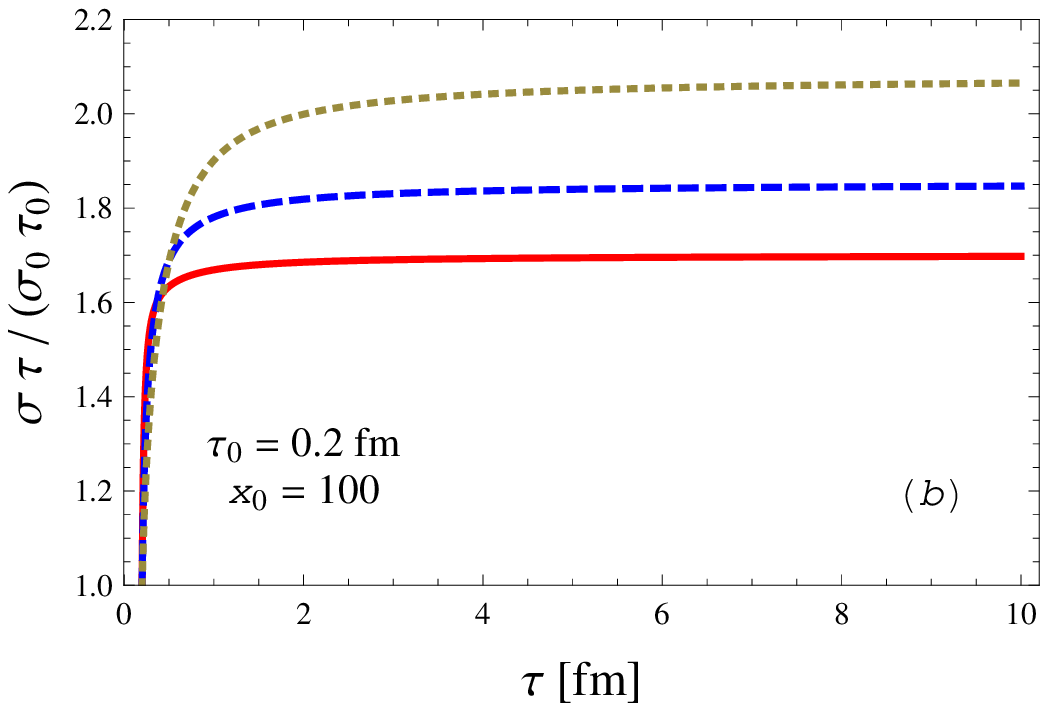}}   \\
\subfigure{\includegraphics[angle=0,width=0.45\textwidth]{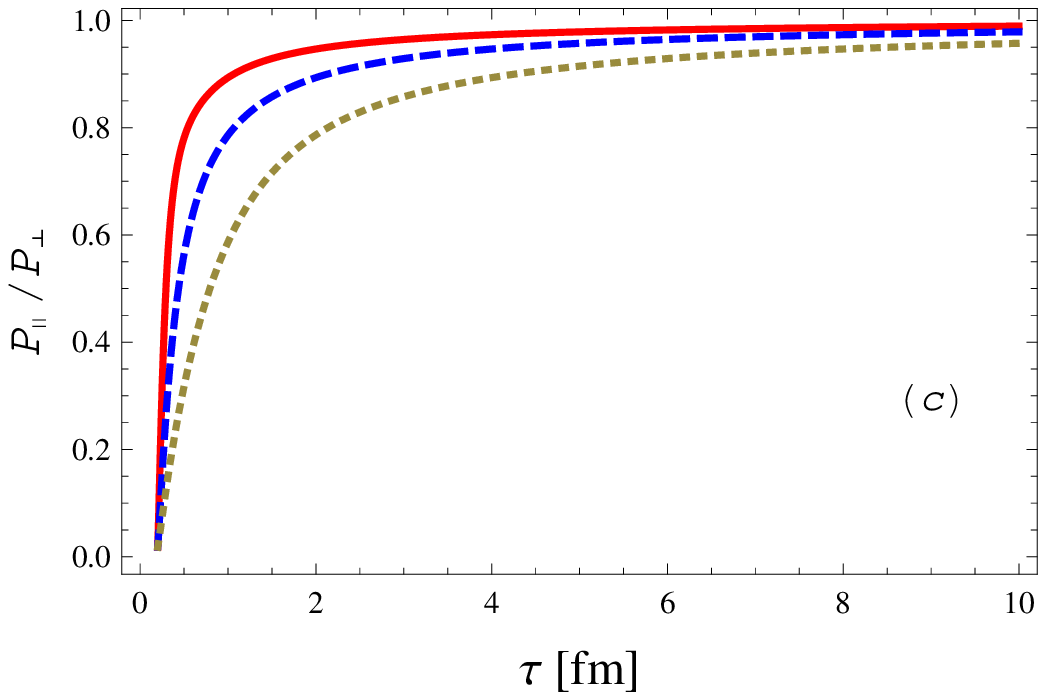}} 
\end{center}
\caption{(Color online) {\bf (a)} The time dependence of the asymmetry parameter $x$ for three different choices of the relaxation time: \mbox{$\tau_{\rm eq}$ =  0.25 fm} (solid line), \mbox{$\tau_{\rm eq}$ =  0.5 fm} (dashed line), and \mbox{$\tau_{\rm eq}$ =  1.0 fm} (dotted line).  {\bf (b)} Entropy density divided by the corresponding values obtained in the Bjorken model. {\bf (c)} Ratio of the longitudinal and transverse pressures shown as a function of the proper time. All results are obtained with the initial asymmetry $x_0=100$.}
\label{fig:XBjPLPT}
\end{figure}

\subsection{Results}
\label{sect:res}

We solve Eq. (\ref{eq12}) numerically with the initial condition $x=x_0$ set at \mbox{$\tau = \tau_0$ = 0.2 fm}. The results of the microscopic models suggest that $P_\parallel \ll P_\perp$ at the early stages of the collisions, thus we first consider the case $x_0=100$. For completeness, in the end of this Section we also show the results obtained with \mbox{$x_0=0.01$}. The time evolution is studied in the time interval: \mbox{0.2 fm $\leq \tau \leq$ 10 fm}. 

In Fig.~\ref{fig:XBjPLPT} we show the time dependence of various physical quantities obtained with $x_0=100$ for three different choices of the relaxation time: \mbox{$\tau_{\rm eq}$ =  0.25 fm} (solid line), \mbox{$\tau_{\rm eq}$ =  0.5 fm} (dashed line), and \mbox{$\tau_{\rm eq}$ =  1.0 fm} (dotted line).

Figure \ref{fig:XBjPLPT} (a) shows the time dependence of the asymmetry parameter $x$.  We observe a fast changes of $x$ in the initial stages of the evolution. Such changes, depending on $\tau_{\rm eq}$, are caused mainly by the fact that $H(x)$ behaves like $6 x^{3/2}$ for large values of $x$. Hence, large initial values of $x$ imply also large (but negative) values of the derivative $dx/d\tau$ at $\tau=\tau_0$. We discuss this behavior in greater detail in Sect. \ref{sect:appA}. 

The behavior shown in  Fig.~\ref{fig:XBjPLPT} (a) indicates also that $x \approx 1$ for $\tau \geq 2 \tau_{\rm eq}$. This is a desired effect showing that the system approaches the local equilibrium state. The way how $x$ approaches unity is described in more detail in Sect. \ref{sect:appB} where the approximate analytic solution is presented. 

\begin{figure}[t]
\begin{center}
\subfigure{\includegraphics[angle=0,width=0.45\textwidth]{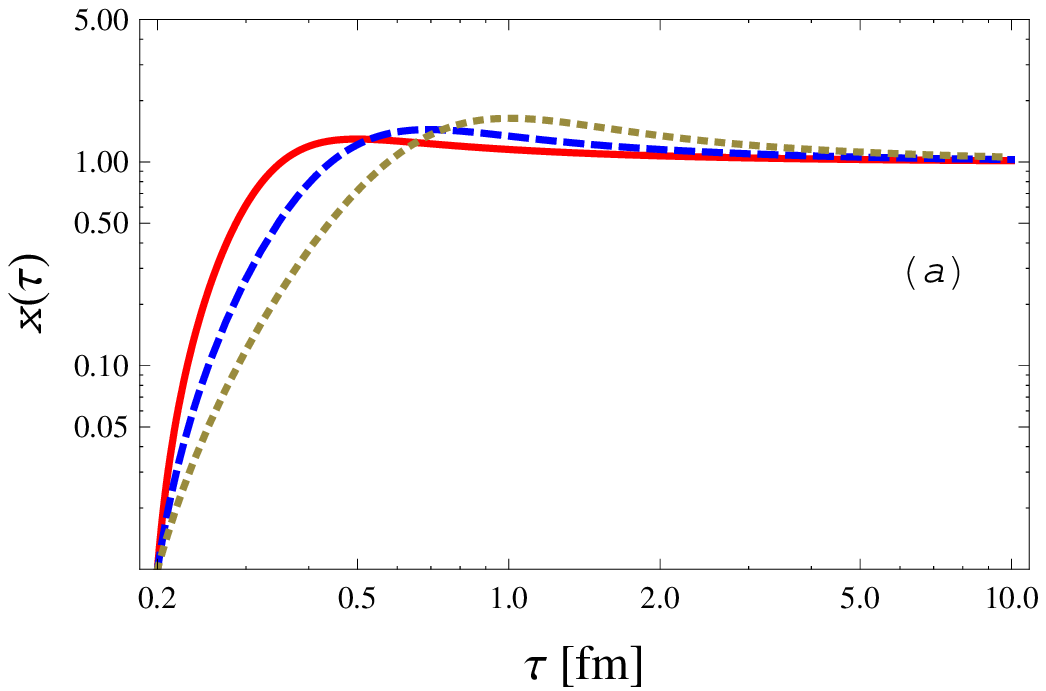}}  \\
\subfigure{\includegraphics[angle=0,width=0.45\textwidth]{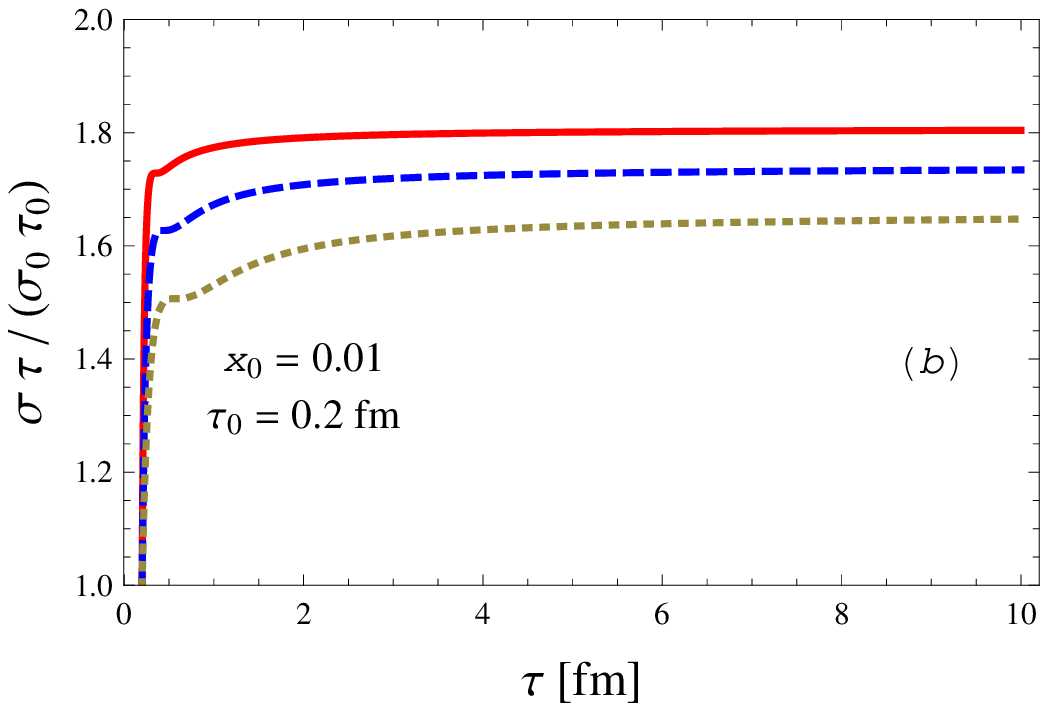}}   \\
\subfigure{\includegraphics[angle=0,width=0.45\textwidth]{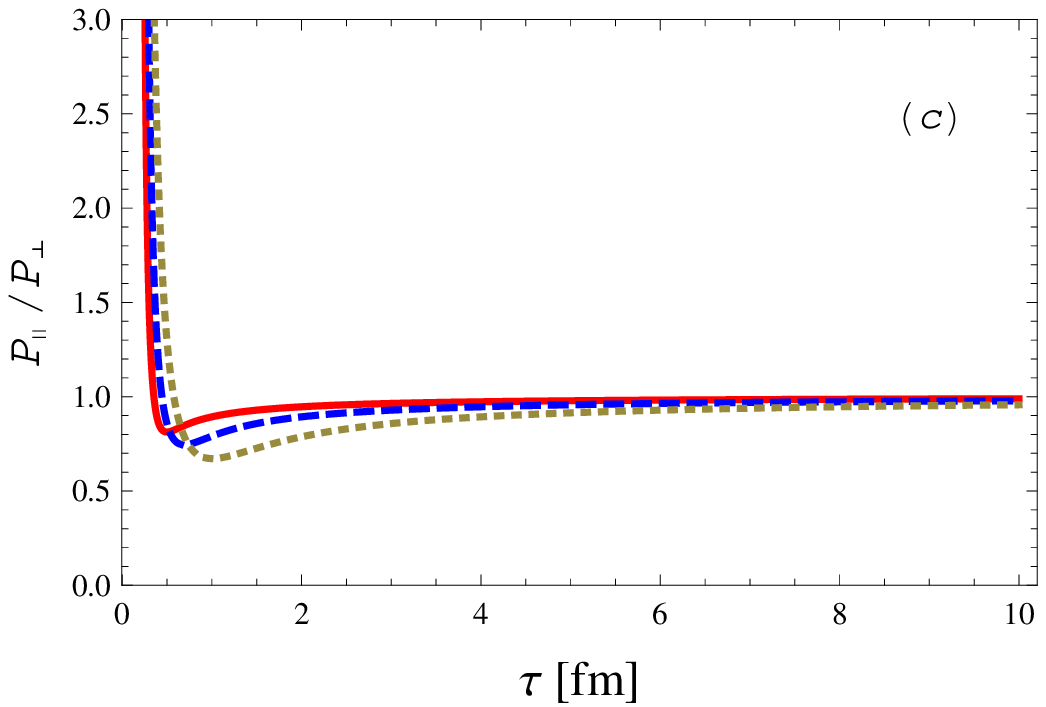}} 
\end{center}
\caption{(Color online) The same as Fig. \ref{fig:XBjPLPT} but for the initial condition $x(\tau_0)=x_0=0.01$.}
\label{fig:XBjPLPT2}
\end{figure}

In Fig.~\ref{fig:XBjPLPT} (b) we compare the time evolution of the entropy density obtained from Eq. (\ref{eq1})  with the Bjorken solution
\begin{equation}
\sigma_{\rm Bj} = \frac{\sigma_0 \tau_0}{\tau}.
\end{equation}
Here $\sigma_0$ is the initial value of the entropy density. We note that the specific value of $\sigma_0$ is irrelevant for our analysis, since the entropy equation is invariant with respect to the multiplication of $\sigma$ by an arbitrary constant. 

The amount of the entropy produced in the regime described by the anisotropic hydrodynamics depends in our case on the relaxation time. For \mbox{$\tau_{\rm eq}$ =  0.25, 0.5, and 1.0 fm} the entropy increases by about 70\%, 85\% and 105\%, respectively. For $\tau \gg \tau_{\rm eq}$ the ratio $(\sigma \tau)/(\sigma_0 \tau_0)$ saturates indicating that the flow attains the form of the Bjorken flow. This  behavior shows again that our framework may be used to model the transition between the highly anisotropic initial behavior and the perfect-fluid stage. 

In Fig.~\ref{fig:XBjPLPT} (c) we show the time dependence of the ratio of the longitudinal and transverse pressure. Again, we show three different time evolutions corresponding to three different relaxation times. For $\tau \gg \tau_{\rm eq}$ the ratio approaches unity and the two pressures become equal. 

Figure \ref{fig:XBjPLPT2} shows the same time evolutions as Fig. \ref{fig:XBjPLPT} but with the initial condition $x_0=0.01$. Our main remark here is that the dynamics of the system governed by Eqs. (\ref{eq1}) and (\ref{eq2}) again leads to the equilibration of the system. The entropy production in the anisotropic phase is similar to the previous case. Interestingly, the time dependence of $x$ and $P_\parallel/P_\perp$ is not monotonic in this case, but anyway $P_\perp \approx P_\parallel$ for sufficiently large evolution times. 
 
\section{Conclusions}
\label{sect:con}

In this paper we have introduced the new framework of highly-anisotropic hydrodynamics with strong
dissipation. The effects of the dissipation are introduced by the special form of the internal entropy source.  The source depends on the pressure anisotropy and vanishes for the isotropic systems to guarantee that the perfect-fluid behavior is reproduced for the locally equilibrated system.

With a simple ansatz for the entropy source satisfying general physical requirements, we have obtained a non-linear equation describing the time evolution of the anisotropy parameter $x$. The non-linearity causes that the initial large (or small) anisotropy parameter approaches asymptotically unity. The rate at which the equilibrium is reached depends on the relaxation-time parameter $\tau_{\rm eq}$. We think that with a suitable chosen value of $\tau_{\rm eq}$, our approach may be useful to model the fast equilibration of matter expected in heavy-ion collisions. In particular, it offers an attractive option for modeling the continuous equilibration of pressures. 

The dynamics of anisotropic fluid determines the changes of the microscopic distribution function $f(x,p)$. Thus, various calculations done so far for the systems in equilibrium may be repeated for the non-equilibrium case. In this way the effects of the non-isotropic dynamics may be analyzed and, more importantly, verified in a very straightforward way.

Our numerical results have been presented for a simple one-dimensional system. Nevertheless, the proposed formalism is general and may be applied to more complicated 2+1 and 3+1 situations. In addition, different forms of the entropy source inspired by different microscopic mechanisms may be analyzed. In this context it is interesting to search for the hints coming from the ADS/CFT correspondence. In our further studies we want to explore such rich possibilities.

\begin{figure}[t]
\begin{center}
\subfigure{\includegraphics[angle=0,width=0.45\textwidth]{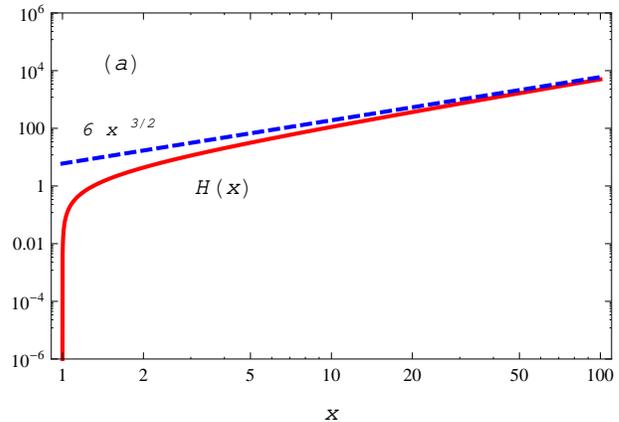}} \\
\subfigure{\includegraphics[angle=0,width=0.45\textwidth]{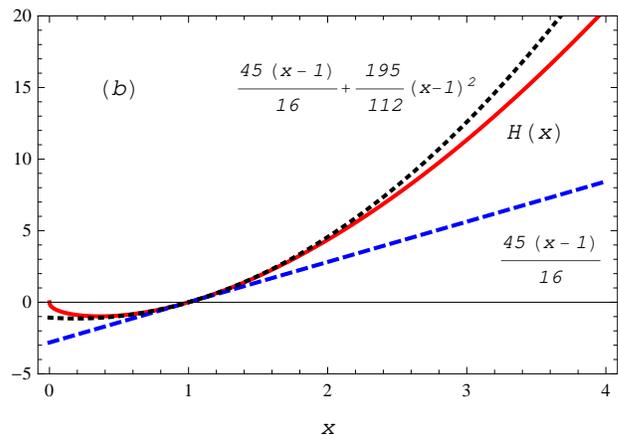}} 
\end{center}
\caption{(Color online) Function $H(x)$ and its approximations in the regions $x \gg 1$ {\bf (a)}  and $|x-1| \ll 1$ {\bf (b)} . }
\label{fig:H}
\end{figure}

\section{Appendices}
\label{sect:app}

\subsection{Analytic solutions for $x \gg  1$}
\label{sect:appA}

For very large arguments $H(x)$ may be approximated by the formula, 
\begin{equation}
H(x) \approx 6 x^{3/2},
\label{inf1}
\end{equation}
see Fig. \ref{fig:H} (a). This leads to the equation
\begin{equation}
 \frac{dx}{d\tau}  =  \frac{2 x}{\tau} - \frac{8  x^{3/2}}{ \tau_{\rm eq}},
\label{inf2}
\end{equation}
which has the following analytic solution
\begin{equation}
x  =  \frac{ \tau_{\rm eq}^2 \tau^2 }
{\left[ 2 ( \tau^2 - \tau_0^2 ) +  \tau_{\rm eq} \tau x_0^{-1/2} \right]^2 }.
\label{inf3}
\end{equation}
We note that large initial values of $x$ imply very large negative values of the derivative $dx/d\tau$, since directly from Eq. (\ref{inf2}) one finds
\begin{equation}
\left. \frac{dx}{d\tau} \right|_{\tau=\tau_0} 
=  \frac{2 x_0}{\tau_0} - \frac{8  x_0^{3/2}}{ \tau_{\rm eq}}.
\label{inf4}
\end{equation}
%

\subsection{Analytic solutions for $|x-1| \ll 1$}
\label{sect:appB}

In the region  $|x-1| \ll 1$ we may use the following approximation, see Fig. \ref{fig:H} (b),
\begin{equation}
H(x) \approx \frac{45}{16}(x - 1) + \frac{195}{112} (x - 1)^2 + \cdots  .
\label{H1}
\end{equation}
If $x$ is close to 1 we take the first term in the series and obtain 
\begin{equation}
 \frac{dx}{d\tau}  =  \frac{2 x}{\tau} - \frac{15}{4\tau_{\rm eq}}(x - 1).
\end{equation}
The solution of this equation has the form
\begin{eqnarray}
x \!\!&=& \!\!
\frac{\tau^2}{\tau_{\rm eq}^2} \exp\left(-\frac{15 \tau}{4 \tau_{\rm eq}}\right)
\left[
A \, \tau_{\rm eq}^2 + \frac{225}{16} \hbox{Ei}\left(\frac{15 \tau}{4 \tau_{\rm eq}}\right)
\right] - \frac{15 \tau}{4 \tau_{\rm eq}}. \nonumber \\
\end{eqnarray}
where  \hbox{Ei}$(x)$ is the exponential integral function and $A$ is an arbitrary integration constant. Using the asymptotic expansion of \hbox{Ei}$(x)$ for $x \gg 1$,
\begin{equation}
\hbox{Ei}\left(\frac{15 \tau}{4 \tau_{\rm eq}}\right) \approx 
\exp\left(\frac{15 \tau}{4 \tau_{\rm eq}}\right) 
\left( \frac{4 \tau_{\rm eq}}{15 \tau} + \frac{16 \tau_{\rm eq}^2}{225 \tau^2} + \cdots \right),
\end{equation}
one obtains
\begin{equation}
 \lim_{\,\,\,\,t \gg \tau_{\rm eq}} x(t) = 1.
\end{equation}
Similar, but much more involved calculations may be done for the case where the second term in the series (\ref{H1}) is included.

\medskip


\end{document}